\documentclass[aps,twocolumn,amsmath,amssymb,floatfix]{revtex4}

\usepackage{graphicx}
\usepackage{dcolumn} 
\usepackage{bm}      
\usepackage{color}

\begin{document}

\title{Rotating helical turbulence. Part I. Global evolution and 
       spectral behavior}
\author{P.D. Mininni$^{1,2}$ and A. Pouquet$^{2,3}$}
\affiliation{$^1$Departamento de F\'\i sica, Facultad de Ciencias Exactas y
         Naturales, Universidad de Buenos Aires and CONICET, Ciudad 
         Universitaria, 1428 Buenos Aires, Argentina. \\
             $^2$Computational and Information Systems Laboratory, NCAR, 
         P.O. Box 3000, Boulder, Colorado 80307-3000, U.S.A. \\
             $^3$Earth and Sun Systems Laboratory, NCAR, P.O. Box 3000, 
         Boulder, Colorado 80307-3000, U.S.A.}
\date{\today}

\begin{abstract}
We present results from two $1536^3$ direct numerical simulations of 
rotating turbulence where both energy and helicity are injected into 
the flow by an external forcing. The dual cascade of energy and helicity 
towards smaller scales observed in isotropic and homogeneous turbulence 
is broken in the presence of rotation, with the development of an 
inverse cascade of energy now coexisting with direct cascades of energy and
helicity. In the direct cascade range, the flux of helicity dominates 
over that of energy at low Rossby number. 
These cascades have several consequences for the 
statistics of the flow. The evolution of global quantities and of the 
energy and helicity spectra is studied, and comparisons with simulations 
at different Reynolds and Rossby numbers at lower resolution are done to 
identify scaling laws.
\end{abstract}
\maketitle

\section{Introduction}

Helicity, the alignment of velocity and vorticity, is a measure of the 
number of links in vorticity field lines, and an indication of lack of 
mirror-symmetry in a flow \cite{Moffatt69}. Isotropic and homogeneous 
turbulence with and without helicity has been thoroughly studied in 
the literature \cite{Borue97,Chen03,Chen03b,Gomez04}. Detailed comparisons 
using direct numerical simulations were carried up to large Reynolds numbers 
and spatial resolutions \cite{Mininni06}. We can think about this problem 
following an analogy by Betchov \cite{Betchov61}: a bag full of nails 
has after being shaken its nails pointing in every direction in space, 
and the resulting spatial distribution is mirror-symmetric. On the other 
hand, a well-shaken bag full of right-handed screws has its screws 
pointing in any direction, but the mirror-symmetry is broken by the 
screws. In isotropic and homogeneous turbulence, the presence of helicity 
does not change properties of turbulence such as the energy spectrum 
\cite{Kraichnan73,Chen03,Gomez04} or the energy decay rate \cite{Lesieur};
 both the energy and the helicity cascade towards 
smaller scales with the same time scales. 

However, the presence of rotation breaks this degeneracy, as the dynamical 
equations for the flow evolution are now sensitive to mirror 
reflections. 
Rotation also breaks down the isotropy of the flow, introducing a 
preferred direction. The development of anisotropies in a rotating flow 
has been studied in great detail \cite{Cambon89,Waleffe93,Cambon97}.
However, the role played by helicity in these flows has been given less 
attention with only a few exceptions \cite{Morinishi01,Galtier03} even 
though the study of helical rotating flows is relevant for many atmospheric 
phenomena \cite{Lilly88,Kerr96,Markowski98}.

In a recent series of papers \cite{Mininni09b,Mininni09}, we presented 
evidence of differences in the scaling laws of rotating flows with 
and without helicity. The presence of helicity in a rotating flow 
changes the energy scaling, as shown in numerical simulations and 
explained with a phenomenological theory \cite{Mininni09}. Changes in 
the directions of the energy and helicity cascades and their associated 
scaling laws have implications for the decay and predictability of a 
helical rotating flow: helical rotating flows decay slower than 
non-helical rotating flows \cite{Teitelbaum09}. These differences imply 
that to model helical rotating flows in nature, subgrid models that take 
into account contributions to the turbulent transport coefficients from 
the helicity are required. In agreement with this, a sub-grid scale model 
based on the eddy-damped quasi-normal Markovian (EDQNM) closure 
\cite{Orszag77,Cambon89,Bellet06} and that takes into account both 
the cascades of energy and of helicity, proved to behave better at 
reproducing simulations of rotating turbulence than models based solely 
on the energy cascade \cite{Baerenzung09}.

However, the grid resolution of these previous simulations of rotating 
helical turbulence (up to $512^3$ grid points) was insufficient to study 
together and at sufficiently high Reynolds number the direct and inverse 
cascades; rather, these cascades were studied separately \cite{Mininni09}. 
Also, the amount by which anisotropies develop at small scales (or the 
possible recovery of isotropy at small scale) was insufficiently quantified. 
In order to 
go further in our analysis of rotating turbulence, we present 
in this paper a detailed study of the results of two direct numerical 
simulations of rotating turbulence at unprecedented resolutions. The 
spatial resolution attained in the simulations allows us to confirm the 
scaling laws for the energy and helicity spectra predicted in 
\cite{Mininni09} in runs where the direct and inverse cascades 
now coexist, each with well defined inertial ranges. We also 
analyze the evolution of global 
quantities in the simulations, the development of anisotropies in the flow, 
as well as scaling laws in the directions parallel and perpendicular to the 
axis of rotation. Section \ref{sec:numeric} describes the simulation and 
the numerical methods used, Sec. \ref{sec:global} describes the time 
evolution of global isotropic and anisotropic norms (energy, helicity, 
dissipation, characteristic length scales), and Sec. \ref{sec:spectral} 
considers the energy and helicity spectra as well as their respective 
fluxes. We also compare with previous runs at different Rossby and Reynolds 
numbers in Sec. \ref{sec:scaling} to identify trends and dependencies with 
the controlling parameters, the Reynolds and Rossby numbers. Finally, 
Sec. \ref{sec:conclusions} gives our conclusions.

In summary, the results presented in this paper provide a detailed description 
of the numerical simulations and compute global and spectral quantities 
often considered in previous studies of rotating turbulence. In a following 
paper (Paper II), we consider in detail the intermittency (or lack thereof) 
of the energy and helicity direct cascades, and study structure functions 
as well as probability density functions of velocity and helicity 
increments. Previous studies of rotating turbulence from numerical 
simulations \cite{Muller07,Mininni09b} showed a substantial decrease 
in the intermittency of the flow when rotation is strong (see also 
\cite{Simand00,Baroud02,Baroud03,Sagaut} for experimental results). The 
present simulations can be used to quantify the impact of helicity in the 
flow intermittency, and the analysis presented in this first paper serves 
as a reference to allow comparisons between these large-resolution 
simulations and previous simulations of rotating turbulence in the 
literature. Overall, these studies allow us to consider the recovery of 
isotropy at small scales, as well as the development of structures at large 
and small scales in the flow, and give a thorough description of rotating 
turbulence at scale separations not considered before in direct numerical 
simulations.

\section{Numerical simulations\label{sec:numeric}}
We solve numerically the equations for an incompressible rotating fluid,
\begin{equation}
\frac{\partial {\bf u}}{\partial t} + \mbox{\boldmath $\omega$} \times
    {\bf u} + 2 \mbox{\boldmath $\Omega$} \times {\bf u}  =
    - \nabla {\cal P} + \nu \nabla^2 {\bf u} + {\bf F} ,
\label{eq:momentum}
\end{equation}
and
\begin{equation}
\nabla \cdot {\bf u} =0 ,
\label{eq:incompressible}
\end{equation}
in a three dimensional box of size $2\pi$ with periodic boundary conditions 
using a parallel pseudospectral code with a spatial resolution of $1536^3$ 
regularly spaced grid points (other resolutions will be briefly considered 
in Sec. \ref{sec:scaling}). Here ${\bf u}$ is the velocity field, 
$\mbox{\boldmath $\omega$} = \nabla \times {\bf u}$ is the vorticity, and 
$\nu$ is the kinematic viscosity. The total pressure ${\cal P}$ modified 
by the centrifugal term  is obtained by taking the divergence of Eq. 
(\ref{eq:momentum}), using the incompressibility condition 
(\ref{eq:incompressible}), and solving the resulting Poisson equation. 
We choose the rotation axis to be in the $z$ direction: 
$\mbox{\boldmath $\Omega$} = \Omega \hat{z}$, with $\Omega$ the rotation 
frequency. Time derivatives are estimated using a second order Runge-Kutta 
method, and the code uses the $2/3$-rule for dealiasing. As a result, 
the maximum wave number is $k_\textrm{max} = N/3$ where $N$ is the linear 
resolution. The code is fully parallelized with the message passing 
interface (MPI) library \cite{Gomez05a,Gomez05b}.

The external mechanical forcing ${\bf F}$ in Eq. (\ref{eq:momentum}) is 
given by a superposition of Arn'old-Beltrami-Childress (ABC) flows 
\cite{Childress},
{\setlength\arraycolsep{2pt}
\begin{eqnarray}
{\bf F} &=& F_0 \sum_{k_F=k_1}^{k_2} \left\{ \left[B \cos(k_F y) + 
    C \sin(k_F z) \right] \hat{x} + \right. {} \nonumber \\
&& {} + \left[C \cos(k_F z) + A \sin(k_F x) \right] \hat{y} + 
   {} \nonumber \\
&& {} + \left. \left[A \cos(k_F x) + B \sin(k_F y) \right] 
   \hat{z} \right\},
\label{eq:ABC}
\end{eqnarray}}
\noindent where $F_0$ is the forcing amplitude, $A=0.9$, $B=1$, $C=1.1$ 
\cite{Archontis03}. An ABC flow, as e.g. in Eq. (\ref{eq:ABC}) for only 
one value of $k_F$, is an eigenfunction of the curl with eigenvalue $k_F$; 
as a result, when used as a forcing function, it injects both energy and 
helicity in the flow. It should be noted that in homogeneous turbulence 
the helicity spectrum cannot develop if it is initially zero (see e.g., 
\cite{Cambon89,Cambon97}), or if an external mechanism does not inject 
helicity. In nature, helicity is created e.g., in the presence of rotation 
and stratification \cite{Moffatt}, or near solid boundaries in rotating 
vessels \cite{Godeferd99}. The use of the ABC forcing, although 
artificial, allows us to study helical rotating turbulence without the 
extra computational cost associated to the presence of boundaries or 
stratification.

Two simulations will be considered in Secs. \ref{sec:global} and 
\ref{sec:spectral}, with the forcing acting from $k_1=7$ to $k_2=8$. This 
leaves some room in spectral space for cascades to develop both at large 
scale and at small scale. The 
viscosity is $\nu = 1.6\times 10^{-4}$ in both runs, and the time step 
$\Delta t = 2.5\times 10^{-4}$. For the first run (run A), $\Omega = 0.06$ 
and it
is started from a flow initially at rest. The run was continued for 
near 10 turnover times, when a turbulent steady state was reached. The 
value of $F_0$ was such that in the steady state, the {\it r.m.s.} velocity 
$U = \left< u^2 \right>^{1/2}$ was of order unity. For the second 
run (run B), $\Omega = 9$, and it is 
 started from the velocity field in run A
at $t\approx 10$. Run B was continued for 30 turnover times. 

The Reynolds, Rossby, and Ekman numbers of the runs quoted in the following 
sections are defined as usual as:
\begin{equation}
\textrm{Re} = \frac{L_F U}{\nu} ,
\end{equation}
\begin{equation}
\textrm{Ro} = \frac{U}{2 \Omega L_F} ,
\end{equation}
and
\begin{equation}
\textrm{Ek} = \frac{\textrm{Ro}}{\textrm{Re}} = \frac{\nu}{2 \Omega L_F^2} ,
\end{equation}
where $L_F = 2\pi/\textrm{min}\{k_F\}$, and the turnover time at the forcing 
scale is then defined as $T = L_F/U$.

In the following, it will be useful to also introduce a micro-Rossby number 
as the ratio of the {\it r.m.s.} vorticity to the background vorticity 
(rotation), 
\begin{equation}
\textrm{Ro}_\omega = \frac{\omega}{2 \Omega} .
\label{eq:microRo}
\end{equation}
The value of the micro-Rossby number plays a central role in the inhibition 
of the energy cascade in rotating turbulence \cite{Cambon97}. If the 
micro-Rossby number is too small, non-linear interactions are completely
damped. According to \cite{Jacquin90}, anisotropies develop in rotating 
flows when the Rossby number $\textrm{Ro}\lesssim 1$ and when the 
micro-Rossby number $\textrm{Ro}_\omega \gtrsim 1$ (it is worth noting that 
the actual values for the transition depend on the particular flow studied).

Based on these definitions, the resulting Reynolds number for both runs 
was $\textrm{Re} \approx 5100$. The Rossby number of run A is 
$\textrm{Ro} \approx 8.5$, while the Rossby number of run B is 
$\textrm{Ro} \approx 0.06$. This results in Ekman numbers 
$\textrm{Ek} \approx 1.6 \times 10^{-3}$ for run A, and 
$\textrm{Ek} \approx 1.1 \times 10^{-5}$ for run B. The micro-Rossby 
number of run B is $\textrm{Ro}_\omega \approx 1.2$ (in all definitions, 
$U$ and $\omega$ are measured in the steady state of run A, or when the 
inverse cascade of energy in run B starts). We therefore study flows with 
large Reynolds numbers, but with moderate Rossby numbers as often 
encountered in geophysical problems. Runs A and B will also be compared 
in Sec. \ref{sec:scaling} with other runs with helical forcing at lower 
resolution as already described in \cite{Mininni09}.

\section{Time evolution\label{sec:global}}
\subsection{Isotropic quantities}

\begin{figure}
\includegraphics[width=8.6cm]{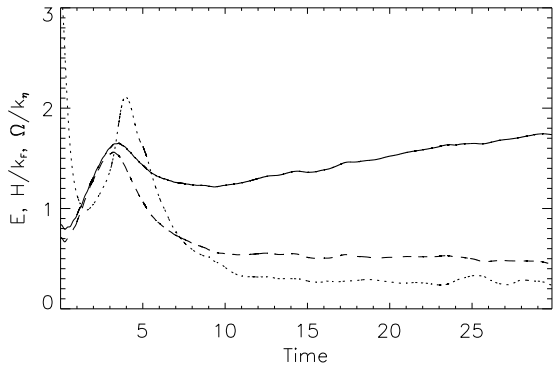}
\caption{Energy (solid), helicity normalized by the forcing wavenumber 
(dashed), and enstrophy rescaled by the dissipation wavenumber $\approx 500$ 
(dotted) as a function of time in run B with $\textrm{Ro}=0.06$.}
\label{fig:enevol} \end{figure}

\begin{figure}
\includegraphics[width=8.6cm]{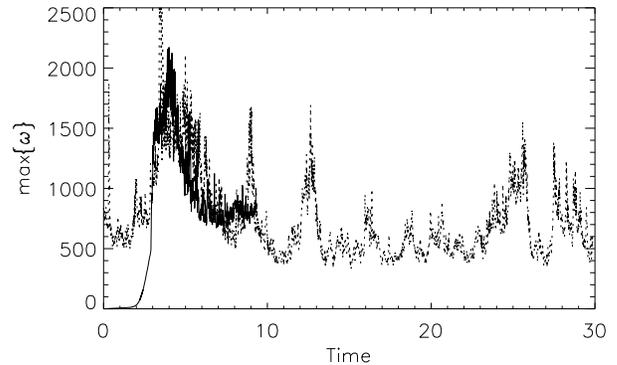}
\caption{Maximum vorticity in the flow as a function of time, for runs A 
(solid) and B (dotted).}
\label{fig:wmaxevol} \end{figure}

Figure \ref{fig:enevol} shows the time evolution of the energy, helicity, 
and enstrophy in run B. After a transient that lasts a few turnover times, 
energy grows monotonically in time. Helicity and enstrophy reach a steady 
state with nearly constant values, smaller than their values at $t=0$ 
which correspond to the steady state in run A.

The enstrophy, proportional to the mean square vorticity, can be used as 
a proxy of the relevance of the rotation near the Taylor scale, as the 
ratio of the {\it r.m.s.} vorticity to the background rotation is 
proportional to the micro-Rossby number. However, turbulence is 
characterized by strong fluctuations of quantities in space and time, and 
one may ask how relevant is the background rotation in the structures that 
correspond to these fluctuations. Another measure of the relevance of 
rotation at the small scales can thus be obtained by looking at regions 
in the flow with maximum vorticity. Fig. \ref{fig:wmaxevol} shows the time 
history of the maximum of vorticity in runs A and B. In run A, which starts 
from a fluid at rest, $\max\{ \omega \}$ grows rapidly from zero and after 
reaching a maximum saturates near $\approx 800$ with strong fluctuations 
around the mean. In run B, started from the last snapshot of run A, 
$\max\{ \omega \}$ starts from the previous value and reaches a maximum 
as the flow becomes anisotropic and the inverse cascade develops, and 
later saturates with a time average value of 
$\max\{ \omega \} \approx 630$ with strong peaks. As a result, in run B 
$\max\{ \omega \}/\Omega \approx 70$.

\begin{figure}
\includegraphics[width=8.6cm]{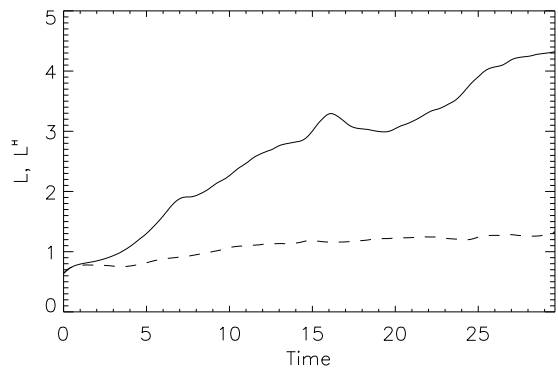}
\caption{Integral scales associated with the energy (solid line) and the 
helicity (dash line) as a function of time in run B.}
\label{fig:levol} \end{figure}

The increase of energy with time observed in Fig. \ref{fig:enevol} is 
associated to an inverse cascade of energy that results in an increase 
in the characteristic size of the energy-containing structures in the flow. 
This is illustrated in Fig. \ref{fig:levol} by the time evolution of the 
isotropic integral scales of the energy 
\begin{equation}
L = \frac{2 \pi}{E} \sum_{k=1}^{k_\textrm{max}} \frac{E(k)}{k} ,
\label{eq:L}
\end{equation}
and of the helicity 
\begin{equation}
L^H = \frac{2 \pi}{H} \sum_{k=1}^{k_\textrm{max}} \frac{H(k)}{k} ,
\label{eq:LH}
\end{equation}
where $E$ and $H$ are respectively the energy and the helicity, and 
$E(k)$ and $H(k)$ are the isotropic energy and helicity spectra. The 
wavenumber $k$ corresponds to the mean radius of the spherical shell 
containing all modes with wave vectors with length between $k-0.5$ and 
$k+0.5$. Initially, both integral scales in run B are close to 
$L_F \approx 0.9$. However, as time evolves the integral scale of the 
energy increases,  while the integral scale of the helicity remains 
approximately constant after a short transient increase.

\subsection{The development of anisotropy}

\begin{figure}
\includegraphics[width=8.6cm]{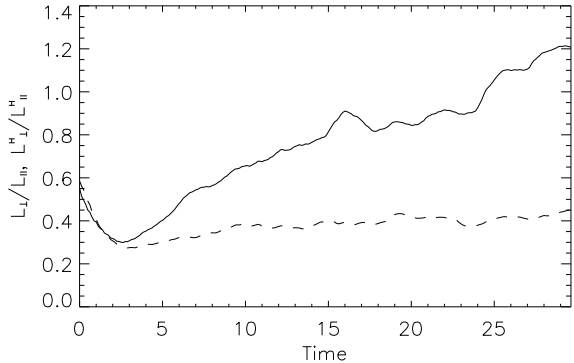}
\caption{Ratio of perpendicular to parallel integral scales, associated 
to the energy (solid) and to the helicity (dashed) as a function of time 
in run B.}
\label{fig:lppevol} \end{figure}

The flow in run B becomes anisotropic after a few turnover times. 
Several indicators can be defined to quantify the development of 
anisotropy. In this section, we will focus on global quantities, and 
later we will consider anisotropy from the point of view of Fourier 
spectra. Directional integral scales as in Eqs. (\ref{eq:L}) and 
(\ref{eq:LH}) can be defined integrating spectra over wave vectors 
parallel and perpendicular to the rotation axis. We will consider the 
energy spectra $E(k_\parallel)$ and $E(k_\perp)$, where $E(k_\parallel)$ 
is obtained by integrating the energy in all modes with wave vectors 
with $k_z=k_\parallel$ between $k_\parallel-0.5$ and $k_\parallel+0.5$ (i.e., 
integrating over planes in spectral space), and $E(k_\perp)$ is 
obtained by integrating the energy in all modes with wave vectors with 
$(k_x^2+k_y^2)^{1/2}$ between $k_\perp-0.5$ and $k_\perp+0.5$ (i.e, 
integrating in cylindrical shells; see e.g., \cite{Alexakis07,Mininni09}). 
These spectra are often referred to in the literature as ``reduced'' 
energy spectra. Similar procedures can be used to construct the reduced 
helicity spectra $H(k_\parallel)$ and $H(k_\perp)$. Then, following 
Eqs. (\ref{eq:L}) and (\ref{eq:LH}), we define perpendicular and 
parallel integral scales
\begin{equation}
L_{\{\perp,\parallel\}} = 2 \pi \frac
    {\sum_{k_{\{\perp,\parallel\}}=1}^{k_\textrm{max}} 
    k_{\{\perp,\parallel\}}^{-1} E\left(k_{\{\perp,\parallel\}}\right)}
    {\sum_{k_{\{\perp,\parallel\}}=1}^{k_\textrm{max}} 
    E\left(k_{\{\perp,\parallel\}}\right)} 
\end{equation}
\begin{equation}
L^H_{\{\perp,\parallel\}} = 2 \pi \frac
    {\sum_{k_{\{\perp,\parallel\}}=1}^{k_\textrm{max}} 
    k_{\{\perp,\parallel\}}^{-1} H\left(k_{\{\perp,\parallel\}}\right)}
    {\sum_{k_{\{\perp,\parallel\}}=1}^{k_\textrm{max}} 
    H\left(k_{\{\perp,\parallel\}}\right)} 
\end{equation}
where the subindex $\{\perp,\parallel\}$ denotes that either parallel or 
perpendicular wave vectors are used.

Figure \ref{fig:lppevol} shows the perpendicular-to-parallel ratio of 
integral scales for the energy and the helicity in run B. As $\Omega$ is 
suddenly increased at $t=0$ from its previous value in run A, these two ratios
first decrease from their initial values. Then, $L^H_\perp/L^H_\parallel$ 
increases slightly and seems to reach a steady state after $t\approx 10$, 
while $L_\perp/L_\parallel$ keeps increasing 
monotonically in time on the whole, 
following the increase in the energy due to the inverse cascade. As a result, the energy
(dominated by large scales) seems to become more anisotropic than the helicity which is concentrated in smaller scales.

\begin{figure}
\includegraphics[width=8.6cm]{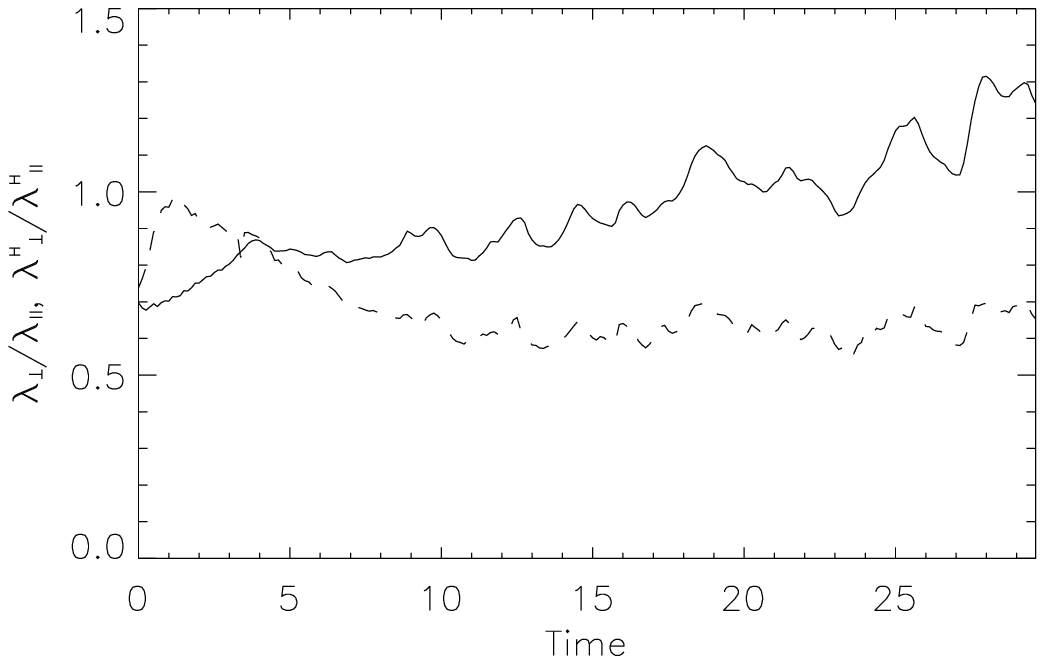}
\caption{Ratio of perpendicular to parallel Taylor scales, associated 
to the energy (solid) and to the helicity (dashed) as a function of time 
in run B at $\textrm{Ro}=0.06$.}
\label{fig:tppevol} \end{figure}

Integral scales are dominated by the contributions from the energy 
containing scales. Note that both $L$ and $L^H$ are, after the short 
transient, larger than the forcing scale $L_F$ (see Fig. \ref{fig:levol}). 
The ratios studied previously then give a global indication of anisotropies 
at scales that are in the inverse cascade range. Anisotropies in the 
(small scale) direct cascade range can be quantified, e.g., by the ratio 
of the perpendicular-to-parallel Taylor scales (Fig. \ref{fig:tppevol}).
The perpendicular and parallel Taylor scales based on the energy and the 
helicity are defined as
\begin{equation}
\lambda_{\{\perp,\parallel\}} = 2 \pi \left[ \frac
    {\sum_{k_{\{\perp,\parallel\}}=1}^{k_\textrm{max}} 
    E\left(k_{\{\perp,\parallel\}}\right)}
    {\sum_{k_{\{\perp,\parallel\}}=1}^{k_\textrm{max}} 
    k_{\{\perp,\parallel\}}^2 E\left(k_{\{\perp,\parallel\}}\right)}
    \right]^{1/2} ,
\end{equation}
\begin{equation}
\lambda^H_{\{\perp,\parallel\}} = 2 \pi \left[ \frac
    {\sum_{k_{\{\perp,\parallel\}}=1}^{k_\textrm{max}} 
    H\left(k_{\{\perp,\parallel\}}\right)}
    {\sum_{k_{\{\perp,\parallel\}}=1}^{k_\textrm{max}} 
    k_{\{\perp,\parallel\}}^2 H\left(k_{\{\perp,\parallel\}}\right)}
    \right]^{1/2} .
\end{equation}

After a transient, the ratio $\lambda^H_\perp/\lambda^H_\parallel$ in 
Fig. \ref{fig:tppevol} stabilizes at its original value at $t=0$, while 
$\lambda_\perp/\lambda_\parallel$ increases with time. However, the 
increase in this ratio is slower than in the case of 
$L_\perp/L_\parallel$, and presents stronger fluctuations in time, 
being associated with scales in the inertial range of the direct energy 
and helicity cascades. Note that, at the onset of the inverse cascade for 
run B, the perpendicular Taylor scale is $\lambda_\perp \approx 0.25$, 
giving for the Taylor Reynolds number of that flow (based on the 
perpendicular scale), 
$\textrm{R}_{\lambda_\perp}=U\lambda_{\perp}/\nu \approx 1600$ 
(for run A, we have $\textrm{R}_{\lambda}\approx 900$ in the turbulent 
steady state; the increase of $\textrm{R}_{\lambda}$ 
in run B is associated 
to the anisotropization and increase of characteristic scales of the 
flow when rotation is increased). Note that these values are larger 
than the values considered in experiments with similar Rossby numbers 
(see e.g., \cite{Baroud02}).

\begin{figure}
\includegraphics[width=8.6cm]{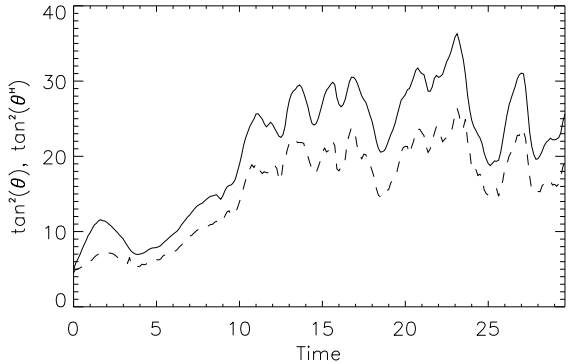}
\caption{Shebalin angles based on the energy (solid) and helicity 
spectra (dashed) as a function of time in run B.}
\label{fig:shebalin} \end{figure}

Yet another measure of small scale spectral anisotropy is given by the 
Shebalin angles \cite{Shebalin83},
\begin{equation}
\tan ^2(\theta) = 2 \frac{\sum_{k_\perp}{k_\perp^2 E(k_\perp)}}
    {\sum_{k_\parallel}{k_\parallel^2 E(k_\parallel)}} ,
\end{equation}
\begin{equation}
\tan ^2(\theta^H) = 2 \frac{\sum_{k_\perp}{k_\perp^2 H(k_\perp)}}
    {\sum_{k_\parallel}{k_\parallel^2 H(k_\parallel)}} ,
\end{equation}
These angles measure the spectral anisotropy level, with the case 
$\tan ^2(\theta) = 2$ corresponding to an isotropic flow. As the previous 
quantities, they only give a global measure of small-scale anisotropy, 
and are a byproduct of axisymmetric energy spectra (see 
\cite{Cambon97,Bellet06}). Figure \ref{fig:shebalin} shows the time 
evolution of the angles based on the energy and on the helicity. The 
helicity at small scales is again more isotropic than the energy. However, 
unlike the previous quantities, the Shebalin angles grow fast and then 
saturate in both cases, reaching a steady state after 10 turnover times.

\begin{figure}
\includegraphics[width=8.6cm]{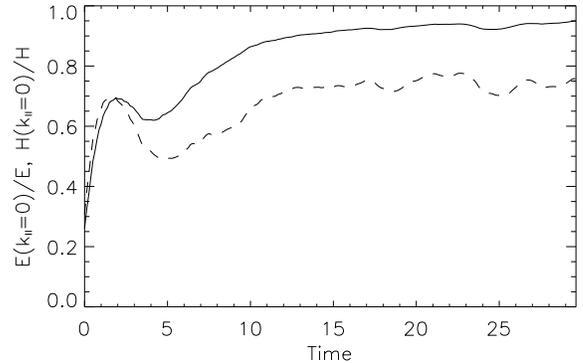}
\caption{Ratios $E(k_\parallel=0)/E$ (solid line) and $H(k_\parallel=0)/H$ 
(dash line) as a function of time in run B.}
\label{fig:anis} \end{figure}

Finally, the amount of energy and helicity in two-dimensional modes 
can be measured with the ratios $E(k_\parallel=0)/E$ and 
$H(k_\parallel=0)/H$ (see Fig. \ref{fig:anis}). Again, the spectral 
distribution of energy is more anisotropic than for helicity. Note 
that at late times a substantial fraction of the energy is in modes 
with $k_\parallel=0$; at $t\approx 29$ near $95\%$ of the energy is 
in those modes, while less than $75\%$ of the helicity is in the 
same modes. All these results indicate that the distribution of energy 
is more anisotropic than that of helicity at all scales. As will be 
discussed next, this is due to the fact that 
helicity only suffers a direct cascade and is therefore 
transported in spectral space to smaller scales which are more isotropic.

\section{Spectral behavior\label{sec:spectral}}

\begin{figure}
\includegraphics[width=8.6cm]{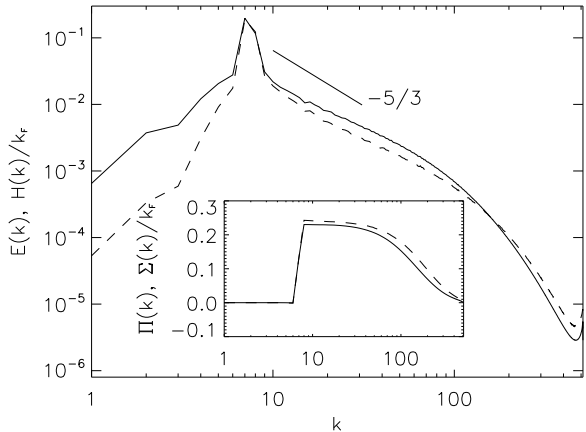}
\caption{Isotropic energy spectrum (solid) and helicity spectrum (dashed) 
normalized by the forcing wavenumber in run A with $\textrm{Ro=8.5}$. Kolmogorov scaling is shown 
as a reference. The inset gives the isotropic energy flux, and the helicity 
flux normalized by the forcing wavenumber.}
\label{fig:spect10} \end{figure}

Figure \ref{fig:spect10} shows the isotropic energy and helicity spectra 
in run A. The run, with negligible rotation effects, displays the usual 
Kolmogorov scaling in the inertial range of the energy and the helicity, 
with a dual cascade of both quantities towards small scales. As in many 
simulations of three-dimensional isotropic and homogeneous turbulence, the 
short inertial range is followed by a bottleneck (which makes the spectra 
slightly shallower) and then by a dissipative range. The dual cascade towards 
smaller scales is further confirmed by examination of the energy and helicity fluxes 
(inset of Fig. \ref{fig:spect10}) which are both positive and constant across 
the inertial range to the right of the forcing wavenumber. At wavenumbers 
smaller than $k_F$, both fluxes are negligible. The small amount of energy 
and helicity observed in the spectra at those wavenumbers is the result of 
backscatter, not of a cascade, and the energy in the large scales displays 
a slope compatible with a $\sim k^2$ scaling (see e.g., 
\cite{Saffman67,Davidson}).

\begin{figure}
\includegraphics[width=8.6cm]{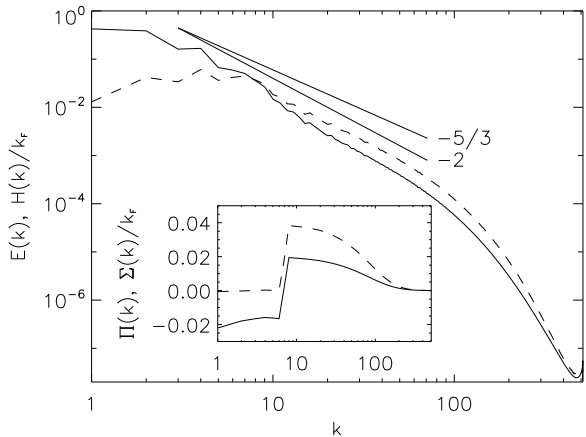}
\caption{Isotropic energy and helicity spectra in run B with $\textrm{Ro}=0.06$ (same labels 
as in Fig. \ref{fig:spect10}); $k^{-5/3}$ (Kolmogorov), and $k^{-2}$ scaling laws are 
shown as a reference. The inset gives the isotropic energy and helicity 
fluxes. Note the excess of (normalized) helicity and of its flux in the small scales.}
\label{fig:spect05} \end{figure}

The energy and helicity spectra and fluxes at late times in run B at $\textrm{Ro}=0.06$ are 
shown in Fig. \ref{fig:spect05}. An inverse cascade of energy develops, as 
evidenced in the spectrum by the pile up of energy at scales larger than 
the forcing, and in the energy flux by a range of wavenumbers with nearly 
constant and negative transfer.
 However, unlike two-dimensional turbulence 
\cite{Boffetta07}, not all the energy injected in the system undergoes an inverse 
cascade: a substantial fraction of the injected energy (approximately 
half at this Rossby number) is still transferred to small scales in a 
direct cascade of energy. Moreover, this direct cascade of energy is 
sub-dominant to a direct cascade of helicity. All the helicity injected 
in the system cascades to small scales, and as a result the helicity flux (properly adimensionalized) is 
larger than the energy flux at all wavenumbers larger than $k_F$.

\begin{figure}
\includegraphics[width=8.6cm]{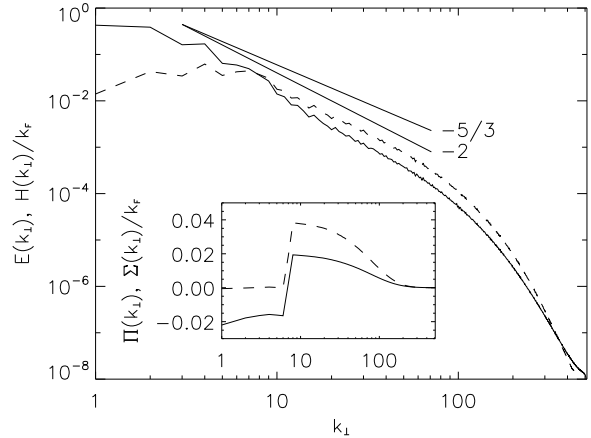}
\caption{Perpendicular energy and helicity spectra in run B with $\textrm{Ro}=0.06$, with same 
labels as in Fig. \ref{fig:spect10}. Kolmogorov and $k^{-2}$ scaling laws 
are shown as a reference. The inset shows the perpendicular energy and 
helicity fluxes.}
\label{fig:sperp05} \end{figure}

Figure \ref{fig:sperp05} displays the perpendicular energy and helicity 
spectra in run B at the same time as in Fig. \ref{fig:spect05}. The 
same features as in the isotropic spectra can be identified, and almost 
no differences are observed between these spectra and the ones showed in 
Fig. \ref{fig:spect05}. This is in agreement with the fact that at late 
times, $\approx 95\%$ of the energy and $\approx 75\%$ of the helicity are 
in modes with $k_\parallel=0$ (i.e., wave vectors perpendicular to the 
axis of rotation). Here and in the isotropic case, the energy spectrum is 
slightly steeper than the helicity spectrum, the result of a dominant 
direct cascade of helicity.

\begin{figure}
\includegraphics[width=8.6cm]{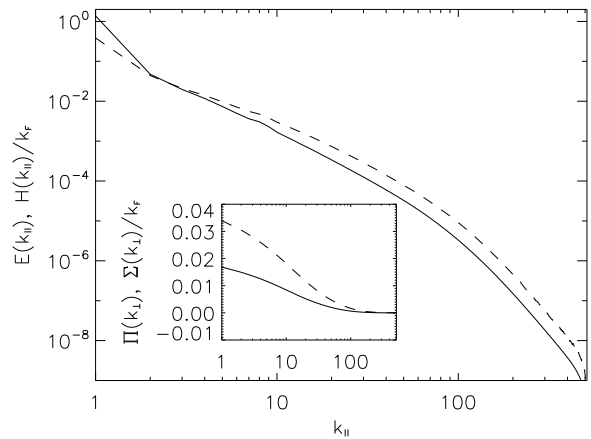}
\caption{Parallel energy and helicity spectra in run B, with same 
labels as in Fig. \ref{fig:spect10}. The inset shows the parallel 
energy and helicity fluxes.}
\label{fig:spara05} \end{figure}

No clear scaling is observed in the parallel spectra (Fig. 
\ref{fig:spara05}). There is an excess of helicity at small scales when 
compared with the energy, as in the previous spectra, but here an 
inertial range cannot be identified. Moreover, the energy and helicity 
fluxes in the parallel wave vectors are positive at all scales, and 
show no approximately constant range. The fluxes peak at 
$k_\parallel = 1$ and decrease fast for larger wavenumbers. This 
indicates that in the parallel direction, a small portion of the energy 
transferred towards large scales by the inverse cascade of energy in 
${\bf k}_\perp$, is transferred back towards small scales although not 
through a cascade. The wavenumber band where forcing  occurs is not 
identifiable either.

\begin{figure}
\includegraphics[width=8.6cm]{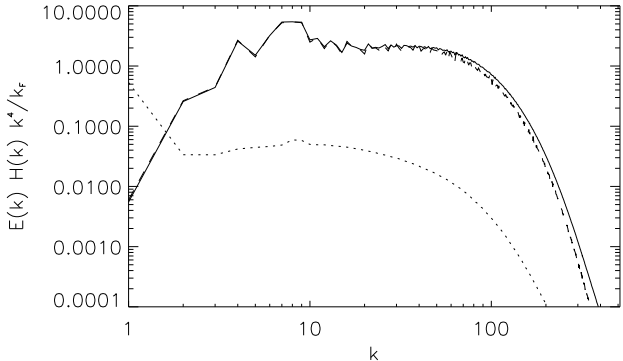}
\caption{Product of the energy and helicity spectra in run B, normalized 
by $k_F$ and compensated by $k^{4}$. The solid line corresponds to the 
isotropic spectra, the dash line to the perpendicular spectra, and the dotted line to 
the parallel spectra.}
\label{fig:rule} \end{figure}

The presence of waves in a rotating flow slows down the direct energy 
cascade, resulting in a steeper energy spectrum than in isotropic and 
homogeneous turbulence. However, unlike isotropic and homogeneous 
turbulence, the slope of the energy spectrum depends on whether the flow 
is helical or not. This is the result of the flow developing, at small 
scales, a dominant direct cascade of energy in the former case, and a 
dominant direct cascade of helicity in the latter. In a flow with 
maximum helicity ($H(k)=kE(k)$), the energy and helicity spectra are, according to these 
arguments (see \cite{Mininni09} for a detailed derivation), predicted to 
be $E \sim k_\perp^{-2.5}$ and $H \sim k_\perp^{-1.5}$. Note that this results in the relative helicity $H(k)/[kE(k)$ independent of wavenumber, i.e. corresponding to an alignment of velocity and vorticity identical at all scales.
 In the general case
(not maximally helical), the energy spectrum gets closer to a
$k_\perp^{-2}$ spectral law, but with both spectra such that their product is still 
$E(k) H(k) \sim k_\perp^{-4}$. The slope of the energy spectrum in the 
direct inertial range of Figs. \ref{fig:spect05} and \ref{fig:sperp05} 
is $\approx 2.1$, while the slope of the helicity spectrum is 
$\approx 1.9$, in good agreement with these predictions (slopes where 
obtained through a least square fit). Figure \ref{fig:rule} shows the 
product of all spectra compensated by $k^{4}$. The isotropic and 
perpendicular spectra show a flat region compatible with this scaling, 
and the results are in good agreement with the arguments based on a 
dominant cascade of helicity in the helical rotating case. No clear 
scaling is observed in the parallel direction.

\section{Scaling with Reynolds and Rossby numbers\label{sec:scaling}}

\begin{figure}
\includegraphics[width=8.6cm]{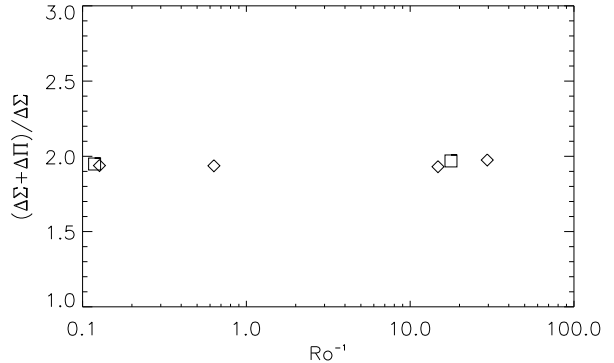}
\caption{Ratio $(\Delta \Sigma + \Delta \Pi)/\Delta \Sigma$ as a function 
of the inverse 
Rossby number for several runs (see text for definition). Squares correspond to runs A and B. 
Diamonds correspond to helical runs with $512^3$ resolution, $k_F=7$, and 
$\textrm{Re} \approx 1200$ for different Rossby numbers, as analyzed in \cite{Mininni09}.}
\label{fig:diff} \end{figure}

In this section we analyze an ensemble of runs in order to see the emergence of scaling laws when examining a range of Reynolds and Rossby numbers; to that effect, we combine runs A and B of this paper 
 with previous simulations at 
different Reynolds and Rossby numbers \cite{Mininni09}. The sub-dominant direct cascade 
of energy in run B is the result of the relation $\tilde \epsilon \sim k_F \epsilon$ 
at the forcing scale (where $\tilde \epsilon$ is the helicity injection rate $dH/dt$ and 
$\epsilon$ the energy injection rate $dE/dt$), together with the development 
of an inverse cascade of energy at small enough Rossby number which 
removes some of the injected energy that would otherwise be available for the direct cascade. Since 
a fraction of the energy injected at $k_F$ goes towards large scales 
(in run B, $|\Pi^-| \approx \epsilon/2$, where $\Pi^-$ is the energy flux towards 
large scales, measured at wavenumbers 
smaller than $k_F$), while most of 
the helicity injected goes towards small scales (with $\Sigma^+$ the 
helicity flux towards small scales), the energy flux towards small scales 
$\Pi^+$ is smaller than $\Sigma^+$ in the entire direct cascade range (see 
e.g., Figs. \ref{fig:spect05} and \ref{fig:sperp05}). This results in a direct cascade dominated by the helicity, 
reminiscent of the pure direct cascade of helicity hypothesized in \cite{Brissaud73}. 
However, here the time scale of the direct cascade is affected by the 
presence of Rossby waves, resulting in the $E(k) H(k) \sim k_\perp^{-4}$ 
rule as discussed in \cite{Mininni09}.

\begin{figure}
\includegraphics[width=8.6cm]{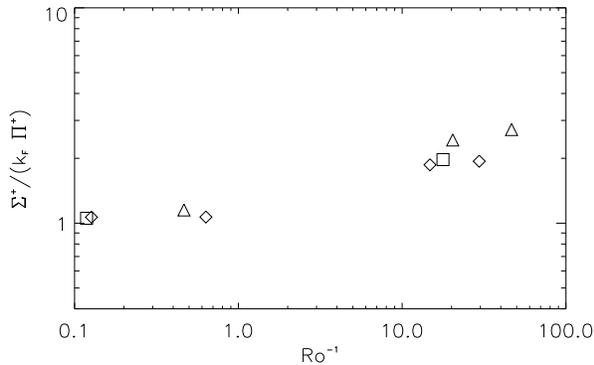}
\caption{Ratio $\Sigma^+/(k_F \Pi^+)$ as a function of the 
inverse Rossby number 
for several runs (see text for definitions). Squares correspond to runs A and B. Diamonds correspond 
to helical runs with $512^3$ resolution, $k_F=7$, and 
$\textrm{Re} \approx 1200$. Triangles correspond to helical runs with 
$512^3$ resolution, $k_F=2$, and $\textrm{Re} \approx 5700$ 
\cite{Mininni09}.}
\label{fig:ratio} \end{figure}

This scenario is mainly based on two hypothesis, that the injection rates 
are related through $\tilde \epsilon \sim k_F \epsilon$ (with the 
proportionality constant equal to one when the forcing injects maximum 
helicity at only one wavenumber, and smaller than one when the forcing 
is not maximally helical), and that the ratio $\Sigma^+/\Pi^+$ in the 
direct cascade range increases with rotation (either monotonically or 
saturating at a small value of $\textrm{Ro}$ such that the direct 
helicity flux becomes dominant). 

That the former condition holds,
in the range of parameters tested in the present study, 
is illustrated in Fig. \ref{fig:diff} which displays 
the normalized ratio $(\Delta \Sigma + \Delta \Pi)/\Delta \Sigma$ as a 
function of the Rossby number for several runs, where 
$$\Delta \Sigma = (\Sigma^+ - \Sigma^-)/k_F = \tilde \epsilon/k_F \ ,$$
 and 
$$\Delta \Pi = \Pi^+ - \Pi^- \ = \ \epsilon\ .$$
Note that these differences (the amount 
of flux to small scales plus the amount of flux to large scales, since 
the latter is negative) are proportional to the helicity and energy 
injection rates $\tilde \epsilon$ and $\epsilon$. For a forcing that 
injects maximum helicity, if $\tilde \epsilon = k_F \epsilon$, then 
$(\Delta \Sigma + \Delta \Pi)/\Delta \Sigma=2$. 
The constancy of $(\Delta \Sigma + \Delta \Pi)/\Delta \Sigma$ for runs 
at different Reynolds and Rossby numbers, as observed in Fig. \ref{fig:diff}, 
builds confidence on the validity of the former hypothesis.

Figure \ref{fig:ratio} shows the ratio $\Sigma^+/(k_F \Pi^+)$, i.e., 
the ratio of energy flux towards small scales to the normalized helicity 
flux towards small scales, as a function of $\textrm{Ro}^{-1}$. While for 
large values of the Rossby number this ratio is close to unity (where 
a dual cascade of energy and helicity takes place, with both quantities 
having the same spectral index in the inertial range), as the Rossby 
number is decreased the ratio becomes larger than one. Simulations with 
very different Reynolds numbers and scale separation between the domain 
size and the forcing scale appear to collapse onto one curve that grows 
as $\textrm{Ro}$ decreases, thus building confidence on the validity of the 
latter hypothesis.

\section{Conclusions\label{sec:conclusions}}

We presented results from two simulations of rotating turbulence at 
large Reynolds numbers and moderate Rossby numbers, with spatial 
resolution of $1536^3$ grid points and forced at intermediate scales, 
allowing for simultaneous direct and inverse cascades of the ideal 
invariants of the flow to develop. The forcing injects both energy and 
helicity, and both inverse and direct cascades of energy develop 
in the case of stronger rotation, 
together with a direct cascade of helicity. The inverse cascade range 
is dominated by the energy, and the flow is anisotropic and only 
weakly helical. On the other hand, in the direct cascade range, the normalized 
helicity flux is larger than the energy flux. The dominance of the 
helicity flux in this range results in the time scales of the direct 
cascade being imposed by the helicity. As a result, the small scale 
helicity spectrum is shallower than the energy spectrum. Both spectra 
scale as $E\approx k_\perp^{-e}$ and $H\approx k_\perp^{-h}$ with 
$e+h=4$ as predicted by phenomenological arguments \cite{Mininni09} 
(for the run with $\textrm{Ro} \approx 0.06$, it was found 
$e\approx 2.1$ and $h\approx 1.9$). The slope of the energy spectrum 
is slightly steeper than what is found in recent numerical simulations 
of non-helical rotating turbulence (see e.g. \cite{Muller07,Mininni09b}).

Comparisons with other simulations of helical rotating flows albeit at 
lower resolution allowed us to build confidence on the dominance of the 
direct cascade of helicity over the energy as the Rossby number is 
decreased. While direct and inverse inertial ranges of energy and 
helicity were identified in the isotropic and perpendicular spectra 
and fluxes, no clear scaling was found in the parallel direction. The 
energy and helicity fluxes in this direction are positive for all 
wavenumbers, and no range of scales with constant flux was found.

The development of anisotropies in the flow was studied using global 
and spectral quantities. In all cases, it was observed that the 
distribution of helicity is more isotropic than the distribution of 
energy. As an example, it was found that at late times $\approx 95\%$ 
of the energy is in modes with $k_\parallel = 0$, while less than 
$75\%$ of the helicity is in the same modes (as a comparision, for run 
A $\approx 24\%$ of both the total energy and helicity are in modes 
with $k_\parallel = 0$). Other measures of anisotropy at both large 
and small scales gave consistent results. The vortical structures that 
develop in such flows and the anisotropy, as well as the recovery 
of isotropy at small scales, 
will be studied in more detail in Paper II using probability density 
functions of velocity and helicity increments, structure functions 
based on the symmetries of the problem 
and three-dimensional visualizations of the velocity and vorticity.

\begin{acknowledgments}
Computer time was provided by NCAR. NCAR is sponsored by the National 
Science Foundation. PDM acknowledges support from grant UBACYT X468/08 
and PICT-2007-02211, and from the Carrera del Investigador Cient\'{\i}fico 
of CONICET.
\end{acknowledgments}


\end{document}